\documentclass[11pt]{article}
\usepackage{amssymb}
\usepackage{amsmath}
\usepackage{amsfonts}
\usepackage{verbatim}
\usepackage{psfrag,graphicx}
\usepackage{amsmath,natbib}
\usepackage{setspace}
\usepackage[left = 1.2in, top = 4in]{geometry}

\newtheorem{theorem}{Theorem}

\newtheorem{lemma}[theorem]{Lemma}

\newenvironment{proof}[1][Proof]{\noindent\textbf{#1.} }{\ \rule{0.5em}{0.5em}}
\oddsidemargin 
\evensidemargin 
\bibpunct{(}{)}{;}{a}{}{,}

\newcommand{\MCMC}{MCMC}
\newcommand{\MH}{MH}
\newcommand{\aimh}{adaptive independent Metropolis-Hastings}

\newcommand{\calZ}{\ensuremath{\cal Z}} 
\newcommand{\calX}{\ensuremath{\cal X}}
\newcommand{\calY}{\ensuremath{\cal Y}}

\newcommand{\norm}{\mid \mid}
\newcommand{\tg}{\widetilde g} 
\newcommand{\ra}{\rightarrow}
\newcommand{\eps}{\epsilon}
\newcommand{\zprime}{z^\prime} 
\newcommand{\xprime}{x^\prime} 
 
\newcommand{\yprime}{y^\prime}
\newcommand{\xiprime}{xi^\prime}
\newcommand{\Kprime}{K^\prime}
\newcommand{\mixt}{\bar g}
\newcommand{\loose}{preliminary} 
\newcommand{\strict}{strict}
\newcommand{\gstar}{\ensuremath{g^\star}}
\newcommand{\tgstar}{\ensuremath{{\tilde g}^\star}}
\newcommand{\var}{\ensuremath{\text{var}}}
\newcommand{\cov}{\ensuremath{\text{cov}}}
\newcommand{\mixtstar}{{\bar g}^*}
\newcommand{\lambdabar}{ {\bar{\lambda}}}
\newcommand{\lambdastar}{{\lambda}^\star}
\begin{document}
\title{Adaptive Independent Metropolis-Hastings by Fast Estimation of
Mixtures of Normals}
\date{\today}
\author{
Paolo Giordani\\
Research Department, 
Sveriges Riksbank\\
paolo.giordani@riksbank.se\\
\and
Robert Kohn\\
Australian School of Business\\
University of New South Wales}
\maketitle

\begin{abstract}
Adaptive Metropolis-Hastings samplers use information obtained from previous
draws to tune the proposal distribution automatically and repeatedly.  
Adaptation needs to be done carefully to ensure convergence to the correct target 
distribution because the resulting chain is not Markovian. We construct
an \aimh{} sampler that uses a mixture of normals as a proposal distribution. 
To take full advantage of the potential of adaptive sampling our algorithm updates 
the mixture of normals frequently,  
starting early in the chain. The algorithm is built for speed and reliability and 
its sampling performance is evaluated with real and simulated examples.  
Our article outlines conditions for adaptive sampling to hold and gives a readily 
accessible proof that under these conditions the sampling scheme generates 
iterates that converge to the target distribution.

\end{abstract}

\setlength{\baselineskip}{14pt} \skip\footins20pt plus4pt minus4pt 
\everydisplay{ \abovedisplayskip
\baselinestretch\abovedisplayskip\belowdisplayskip
\abovedisplayskip\abovedisplayshortskip
\baselinestretch\abovedisplayshortskip\belowdisplayshortskip
\baselinestretch\belowdisplayshortskip}

\begin{doublespace}
\section{Introduction} \label{s:introduction}

Bayesian methods using Markov chain Monte Carlo (\MCMC) 
simulation have greatly influenced statistical and 
econometric practice over the past fifteen years because of their 
ability to estimate complex models and produce
finite sample inference. A key component in implementing \MCMC{} 
simulation is the Metropolis-Hastings (\MH) method 
\citep{metropolis53,hastings70}, which requires the
specification of one or more proposal distributions. The speed 
at which the chain converges to the posterior  distribution 
and its ability to move
efficiently across the state space depend crucially on whether the proposal 
distribution(s) provide good approximations to the target distributions, either 
locally or globally. Given the key role played by proposal 
distributions, it is natural to use experience from preliminary runs to {\em tune} 
or {\em adapt} the proposal to the target. 
We define {\em \strict{}} 
adaptation as adaptation that is subject to theoretical 
rules which ensure that the iterates converge to realizations from the 
correct target (posterior) distribution. All other adaptation of the 
\MH{} kernel will be called {\em \loose} adaptation,  
whose purpose  is  either to obtain an 
adequate proposal density before switching to non-adaptive \MCMC{} 
sampling, or as the starting point for strict adaptive sampling.  

Despite the different theoretical requirements of \loose{} and 
\strict{} adaptation, a great deal of care is needed in 
constructing both types of adaptive sampling schemes because 
the ultimate goal is to obtain reliable estimates of functionals 
of the target distribution as quickly as possible. 

The literature on adaptive MCMC{} methods follows two main strands.
Adaptation by \textit{regeneration} stems from the work of \cite{gilks98}.   
Our article focuses exclusively on \textit{diminishing
adaptation} schemes. Important theoretical advances in diminishing
adaptation were made by \cite{holden98}, 
\cite{haario01}, 
\cite{andrieu01}, 
\cite{andrieu06}, 
\cite{andrieu05}, 
\cite{atchade05}, 
\cite{nott05} and \cite{roberts06}.  
The proofs of convergence for strict adaptive sampling
 are more complex than for the non adaptive 
case as the iterates are not Markovian because the \MH{} kernel can depend 
on the entire history of the draws. 
Although more theoretical work on adaptive sampling can be expected, the existing 
body of results provides sufficient justification and guidelines to build adaptive 
\MH{} samplers for challenging problems. 

Research is now needed on how to design efficient and reliable adaptive samplers for broad classes of problems. This more applied literature 
mostly focuses on random walk Metropolis, see for example 
\cite{roberts06_applied}. 
Partial exceptions are \cite{gasemyr03} 
who uses normal proposals for an independent Metropolis-Hastings, but limits 
the tuning of the parameters to the burn-in period, and \cite{hastie05},  
who mixes random walk and independent \MH{} in reversible jump problems, in 
what we call two step adaptation in section~\ref{s:twostep}.  Independent \MH{}
schemes are implemented by \cite{nott05} to sample discrete state
spaces in variable selection problems (e.g. to learn if a variable is in or
out), and by \cite{giordani08} 
to learn about interventions, such as breaks or outliers, in dynamic mixture models. 

Our paper contributes to the development of both \loose{} and \strict{} 
algorithms for adaptive independent \MH{}  sampling in continuous state spaces. 
Increased sampling efficiency is obviously one important goal, 
particularly in cases where current best practice (typically some version of 
random walk Metropolis or Gibbs sampling) is less than 
satisfactory. But equally important achievements of adaptive 
schemes may be to expand the set of
problems that can be handled efficiently by general purpose 
samplers and to
reduce coding effort. In particular, adaptive schemes can reduce 
dependence on the use of conjugate priors. Such priors make it  
easier to implement MCMC schemes, 
but are less necessary when using adaptive sampling. 

Our adaptive sampling approach is built on four main ideas. 
The first is to combine \loose{} 
and strict adaptation into one estimation procedure. The second is to estimate 
mixtures of Gaussians from the history of the draws and use them as proposal 
distributions for independent \MH{} in both parts of the 
adaptation. The third is to perform this estimation frequently,
particularly during the \loose{} 
adaptation stage, 
a strategy we call \textit{intensive 
adaptation}. The fourth is 
to ensure that the theoretical conditions for the correct  
ergodic behavior of the sampler are respected during  
\strict{} adaptation. To apply
these ideas successfully, estimation of the mixture parameters 
needs to be fast, reliable, and robust. We achieve a good 
balance of these goals by carefully selecting and tailoring to 
our needs algorithms developed in the clustering literature.

We study the performance of our adaptive sampler  in three 
examples in which commonly used Gibbs schemes can be very inefficient
and compare it with an adaptive random walk Metropolis sampler proposed by 
\cite{roberts06_applied} that builds on the work .                  

Our paper also provides conditions and outlines a proof that our 
\strict{} adaptive sampling scheme converges to the correct 
target distribution and gives convergence rates. 

\section{Two-step adaptation and intensive 
adaptation\label{s:twostep}}
A necessary condition for a successful \aimh{} (AIMH)
sampler is that, given a sizable
sample drawn from the target $\pi (z)$, the suggested algorithm can build a
proposal $q(z)$ which is sufficiently close to the target for IMH to perform
adequately. A two-step adaptive strategy is also feasible whenever 
the answer is positive. We loosely define \textit{two-step adaptation} as a 
sampling scheme in which a rather thorough exploration of the target 
density is carried out in the first part of the chain by a 
sampler other than IMH (such as random walk Metropolis) before switching to a more 
efficient IMH sampler with proposal density tuned on the first-stage 
draws. An early version of such a two-step procedure is proposed by 
\cite{gelman92}. \cite{hastie05} provides an interesting application to 
reversible jump problems. 

Two-step adaptation is relatively simple and safe and in some 
cases is capable of achieving sizable efficiency gains. However, it has the following limitations. First, if the first stage sampler fails to adequately
explore a region of the state space, the proposal built for the second 
stage will also inadequately cover that region. To 
reduce this risk we may need a very large number of iterations in the 
first phase, which may be particularly time consuming 
if the first stage sampler is inefficient. 
Second, there may be no saving of coding effort if 
the first stage 
sampler generates from several conditional distributions, as 
in Gibbs or Metropolis-within-Gibbs, in order to be efficient. 

We loosely define \textit{intensive adaptation} as an AIMH scheme in which
the proposal distribution is updated frequently, particularly in the early
part of the chain. Building a sequence of increasingly good proposal
densities in intensive adaptation is more demanding than building a
proposal density once based on thousands of draws. The question is
whether we can adequately explore the target distributions given an initial
proposal $g_{0}(z)$ but no draws. The answer inevitably depends on the
initial proposal $g_{0}(z)$, on the target $\pi (z)$, and on the details of
the sampling scheme. However, it is possible to outline some general dangers
and opportunities offered by intensive adaptation.

Among the advantages, if the proposal distribution is sufficiently flexible,
frequent tuning of its parameters and continuing adaptation for the entire
length of the chain reduces the risk of a long run of rejections 
and increases the chances of good performance when the initial 
proposal approximates the target poorly.

Estimating proposal densities based on a small number of draws presents some
dangers that the designer of an AIMH scheme should try to minimize. For
example, suppose that we predetermine the iteration, say $j$, at which the
proposal is first updated. If the first $j$ draws have all been rejected,
then a proposal distribution based entirely on these draws is degenerate and
makes the chain reducible. If too few draws have been accepted, the proposal
may be very poor.  We employ the following strategy to prevent these outcomes.
First, we initially update the proposal at a
predetermined number of \emph{accepted} draws. Second, after fitting a
mixture of normal distributions to past draws, we fatten and stretch its
tails by creating a mixture of mixtures as described in section~\ref{Section: 
Design}.  Third, we let the proposal be a mixture where one component is the
initial proposal $g_{0}(z)$, which should of course have long tails. This is
similar to the \textit{defensive mixtures} approach studied by 
\cite{hester98} 
for importance sampling. The same provisions reduce the risk of
adapting too quickly to a local mode.

\textheight 9.2in
\section{Some theory for adaptive sampling} \label{S: theory}
\label{Section: Design}
A \textit{diminishing adaptation} \MH{} sampler performs the 
accept/reject step
like a standard \MH{}, but updates the proposal distribution  
using the history of the draws. This updating is `diminishing' 
in the sense that the proposal
distribution settles down asymptotically (in the number of 
iterations).

This section outlines the theoretical framework for 
\strict{} \aimh{} sampling as used in our article that gives 
some support for our practice. 
The appendix outlines proofs of the theoretical 
results, which extend similar results in \cite{nott05} for the 
case of a finite state space. Our aim is to provide simple 
accessible proofs that will help to popularize the adaptive 
methodology. All densities in this section are with respect to Lebesgue 
measure or counting measure, which we denote as $\mu\{\cdot\}$. 

Let \calZ{}  be a sample space and $\pi(z)$ a target density on 
\calZ. We  use the following adaptive \MH{} scheme to construct a sequence 
of random variables $\{ Z_n, n \geq 2 \}$  whose distribution 
converges to $\pi(z)$. We assume that $Z_0$ and $Z_1$ are 
generated independently from some initial density $q_I(z)$. 
In our examples, this is a heavy tailed version of the Laplace 
approximation to the
posterior distribution. For $n \geq 1 $, let $q_n(z; \lambda_n)$ 
be a proposal
density for generating  $Z_{n+1}$, where $\lambda_n$ is a 
parameter vector that is 
based on  $Z_0 = z_0, \dots,     Z_{n-1}=z_{n-1}$. Thus, given 
$Z_n = z$, we 
generate  $Z_{n+1} =  z^\prime $ from $q_n$, and then with probability 
\begin{align} \label{e:alpha} 
\alpha_n(z,z^\prime) & = \min \biggl(  1, \frac{\pi(\zprime)}{ \pi(z)} 
\frac{q_n(z; \lambda_n) }{q_n(\zprime;\lambda_n) } \biggr ) 
\end{align}
we take $Z_{n+1} = \zprime$; otherwise we take $Z_{n+1} = z$. 
Our choice of $q_n(z; \lambda_n)$ is 
\begin{align}\label{e:proposal} 
q_n(z;\lambda_n) & = \omega_1 g_0(z) + (1-\omega_1) \mixt_n(z;\lambdabar_n) 
\end{align}
where $0 <  \omega_1 <1 $ and the density $g_0(z)$ does not depend on $\lambda_n$. We usually fix $\omega_1$ so that $\lambda_n = \lambdabar_n$; otherwise $\lambda_n = (\lambdabar_n, \omega_1)$. The density $\mixt_n(z; \lambdabar_n)$ is an approximation to $\pi(z)$ whose form is 
discussed below, in section~\ref{s:clustering}  and in appendix~1. 
We assume that there exists a constant $K > 0$ such that for all 
$z \in \calZ$ 
\begin{align}\label{e:domin}
\frac{\pi(z)}{g_0(z)} \ \leq \ K,  \quad \frac{\mixt_n(z; \lambdabar_n)}{g_0(z) } \ \leq \ K\ , 
\quad \text{and} \quad \frac{q_I(z)}{g_0(z)} \ \leq \ K
\end{align}
and 
\begin{align} \label{e:dimn adap} 
\sup_{z \in \calZ} \biggl | \biggl  ( \mixt_n(z; \lambdabar_n) - \mixt_{n+1} (z; \lambdabar_{n+1})  \biggr )/g_0(z) \biggr | 
& = a_n 
\end{align}
where $a_n = O(n^{-r})$ for some $r > 0 $ 
almost surely. If $\calZ$ is compact then \eqref{e:domin} holds almost 
automatically. If, in addition, $\lambdabar_n$ is based on means and covariances of 
the $z$ iterates then we can show that $\norm \lambdabar_n - \lambdabar_{n+1}\norm = 
O(n^{-1})$ and equation \eqref{e:dimn adap} 
also holds. In relation to \eqref{e:domin}, we note that in the non-adaptive case, that is $q_n(z;\lambda_n) = q(z)$ for all $n$, \cite{mengersen96} show that $\pi(z)/q(z) \leq K$ for all $z$ is a necessary and sufficient condition for geometric ergodicity. 

Under conditions \eqref{e:domin} and \eqref{e:dimn adap}, the following results 
are proved in Appendix~2. 

\begin{theorem} \label{thm:theorem1} 
For all measurable subsets $A$
\begin{align} \label{e:theorem1} 
\sup_{A \subset \calZ} \mid \Pr(Z_n \in A) - \pi(A) \mid & \ra 0 \quad \text{as} \quad n \ra \infty.
\end{align}
\end{theorem}

\begin{theorem} \label{thm:theorem2} 
Suppose that $h(z)$ is a measurable function that is square integrable with respect to the density $g_0$. Then, almost surely, 
\begin{align} \label{e:theorem2} 
\frac{1}{n} \sum_{j=1}^n h (Z_j) \ra E_\pi(h(Z)) \quad  \text{as   } \quad n \ra 
\infty.  
\end{align}
\end{theorem}

We now describe the form of $\mixt_n(z;\lambdabar_n)$ in 
\eqref{e:proposal}. For conciseness, we shall often omit showing dependence on $\lambda$; e.g. we will write $\mixt_n(z;\lambdabar_n)$ as $\mixt_n(z)$. 
Let $\gstar_n(z) = \gstar_n(z; \lambdastar_n)$ be a mixture of normals obtained using 
k-harmonic means clustering as described in 
section~\ref{s:clustering} and appendix~1, and let $\tgstar_n(z)$ be a second 
mixture of normals having the same 
component weights and means as $\gstar_n(z)$, but with its 
component variances inflated by a factor $k>1$. Let 
\begin{align} \label{e:fatten} 
\mixtstar_n(z) & = 
\omega_2^\prime \tgstar_n(z) + (1-\omega_2^\prime) \gstar_n(z) \ , 
\end{align}
where $\omega_2^\prime = \omega_2/(1-\omega_1)$ with $\omega_1 $ defined in 
\eqref{e:proposal}, $\omega_1 > 0 , \omega_2 > 0 $, and $\omega_1 + \omega_2 < 1$. 
We note that $\mixtstar_n(z)$ is also a mixture of normals, and we say that 
$\mixtstar_n(z)$ is obtained by \lq stretching and fattening\rq{}  the tails of 
$\gstar_n(z)$. This strategy for obtaining heavier tailed mixtures is used 
extensively in our paper. 

To allow for diminishing adaptation, we introduce the sequence $\beta_n$, where $0 \leq \beta_n \leq 1$, and define 
\begin{align} \label{e:proposal1} 
\mixt_n(z) & = (1-\beta_n) \mixtstar_n(z) + \beta_n\mixt_{n-1} (z) 
\end{align}
with $\mixt_n(z)= \mixt_n(z; {\lambdabar}_n)$, where ${\lambdabar}_n$ is a function of $\lambdastar_n$ and $\beta_n$. Alternatively, we can define 
\begin{align}\label{e:proposal1a} 
\lambdabar_n &= (1-\beta_n)\lambda_n^\star + \beta_n \lambdabar_{n-1}
\end{align}
and $\mixt_n(z)=\mixt_n(z; {\lambdabar}_n)$. 

If we restrict $\calZ $ to be compact and let $\beta_n \ra 1$ at an appropriate rate
then it is straightforward 
to check in most cases that the dominance  and diminishing adaptation conditions, 
\eqref{e:domin} and \eqref{e:dimn adap}, hold. If $\calZ$ is unconstrained but we restrict 
the $\lambda_n^\star$ to lie in a bounded set then we can do rough empirical checks that
\eqref{e:domin} and \eqref{e:dimn adap} hold by taking $g_0(z)$ to be sufficiently heavy 
tailed. In our empirical examples we often find that we can take $\beta_n = 0 $ for all $n$ because $\lambda_n^\star $ converges to a $\lambda^\star$ at a sufficiently fast rate  as $n$ increases. This means that if $g_0(z)$ is sufficiently heavy tailed  then \eqref{e:domin} and \eqref{e:dimn adap} hold as $n$ increases. 

Section~7 and appendix~F of \cite{andrieu06} give general convergence results for 
\aimh{} and \cite{roberts06} give an elegant proof of the 
convergence of adaptive sampling schemes. However, we believe that readers may find the 
conditions \eqref{e:domin} and \eqref{e:dimn adap} and the proofs of 
Theorems~\ref{thm:theorem1} and 
\ref{thm:theorem2} easier to understand for the methodology proposed in our article.

\section{Implementation of the adaptive sampling scheme} \label{s:implementation}
This section outlines how our sampling scheme is implemented. We 
anticipate that readers will use this as a basis for their own 
experimentation. 
In the \loose{} phase the density $g_0(z)$ in \eqref{e:proposal} is a mixture 
of the Laplace approximation to the posterior and 
a heavier tailed version of the Laplace approximation, using weights of 0.6 
and 0.4. By a heavier tailed version we mean a distribution with the same mean 
but 
with a covariance matrix that is 25 times larger. If the Laplace approximation 
is unavailable, then we use the prior. At the 
end of the \loose{} phase, $g_0(z)$ is 
constructed as a mixture of the last estimated mixture, which we call 
$g_{\text{last}}(z)$ say, and a heavier tailed version of 
$g_{\text{last}}(z)$. That is, the component weights and means are set to 
those of $g_{\text{last}}(z)$, 
and its component variances equal to $k = 25$ times the 
component variances of $g_{\text{last}}(z)$. 

In our empirical work we set the 
weights  $\omega _{1}=0.05,$ and $\omega _{2}=0.15$ in \eqref{e:proposal} and 
\eqref{e:fatten}. We also inflate the component variances of $g_n^*(z)$ by a 
factor of $k = 16$ to obtain the corresponding variances of $\tg_n^*(z)$. 
These values have been found to work well but are not optimal in any specific 
sense. 
We conjecture that the speed of convergence and efficiency of our sampler 
could be
further improved with a more careful (and possibly adaptive) choice of these
parameters. Any other value of $k$ in the range $9$-$25$ and of $\omega _{1}$
and $\omega _{2}$ in the range $0.05-0.3$ worked well for the examples given in
this paper, and $\omega _{1}$ could be set to 0 in the \loose{} phase without 
affecting the results.

In our empirical work, during the \loose{} phase, when there are 2 to 4 unknown 
parameters as in the inflation and stochastic volatility examples, we first 
re-estimate the k-harmonic means mixture after 20 {\em accepted} draws in order to 
ensure that the 
estimated covariance matrices are positive definite.  If our parameter space is bigger then we would 
increase that number appropriately. We then re-estimate the mixture after 50, 
100, \dots, 350, 400, 500, \dots, 1000, 1500, \dots , 3000 draws and then every 
1000 draws thereafter.  We also recommend updating the proposal following a period 
of low acceptance probabilities in the MH step. Specifically,  
we re-estimate the 
mixture parameters if the
average acceptance probability in the last $L$ iterations is lower than $%
\alpha _{L}$, where we set $\alpha _{L}=0.1$ and $L=10$. Low acceptance
probabilities signal a poor fit of the proposal, and it is therefore
sensible to update the proposal to give it a better chance of covering the
area around the current parameter value. Since it is unclear that this
does not violate any of the conditions required for the ergodicity of our
adaptive sampler, we limit the updating of the proposal at endogenously
chosen points to a preliminary period, after which the proposal is updated
only at predetermined intervals. The end of the \loose{} adaptation 
period could be set ex-ante, but we prefer to determine it endogenously by 
requiring the smallest acceptance probability in the last $M$ iterations to be 
higher than $\alpha _{M},$ where $M$ is set to 500 and $\alpha _{M}$ to 0.02. 
During the period of \strict{} adaptation, we update the proposal every 1000 
draws. 

We conjecture that Theorems~1 and 2 will still hold if we update the proposal 
after a sequence of low acceptance probabilities so that we could also use this 
update strategy during the period of \strict{} adaptation. However, we have not implemented
this in our empirical analyses. 

The estimation of the mixture of normals can become slow when the number of
iterations is large. To avoid this problem, after $1000$ accepted draws we
only add every $j$-$th$ draw to the sample used to estimate the mixture
parameters, where $j$ is chosen so that the mixture is not estimated 
on more than $10000$ observations.

When most parameters are nearly normally distributed, fitting a mixture of
normals to all the parameters is problematic in the sense that the chances
of finding a local mode with all parameters normally distributed is quite
high (though depending on the starting value of course). This is true of
clustering algorithms and also of EM-based algorithms. To improve the
performance of the sampler in these situations, we divide the parameter
vector $\theta $ into two groups, $\theta _{1}$ and $\theta _{2}$, where
parameters in $\theta _{1}$ are classified as approximately normal and
parameters in $\theta _{2}$ are skewed.\footnote{%
Our rule of thumb is to place a parameter in the `normal' group if its
marginal distribution has $|s|<0.2,$ where $s$ is the skeweness. Our
fattening the tails of the mixture should handle the kurtosis, though this
would optimally be done with mixtures of more flexible distributions than
the normal.} A normal is then fit to the first group and a mixture of $p$
normals to the second. For $\theta _{1}$, we can compute the mean $\mu
_{\theta _{1}}$ and covariance matrix $\Sigma _{\theta _{1}}$\ inexpensively
from the draws. For $\theta _{2},$\ we fit a mixture of normals as detailed
below, estimating\ probabilities $\pi _{1},...,\pi _{p},$\ means $\mu
_{1},...,\mu _{p},$\ and covariance matrices $\Sigma _{1},...,\Sigma _{p}.$\
We then need to build a mixture for $\theta =\{\theta _{1},\theta _{2}\}$.
The mean is straightforward: for the normal parameters, all components have
the same mean. The diagonal blocks of the covariance matrices $\Omega _{i}$
corresponding to $\var(\theta _{1})$ and $\var(\theta _{2})$ for component $i$
are also straightforward. The off-diagonal blocks of $\Omega _{i},$
corresponding to $\cov(\theta _{1},\theta _{2})$ is computed as\textit{\ }%
\begin{equation*}
\Omega _{i}^{12}=\sum_{t=1}^{n}\pi _{i,t}^{\ast }[(\theta _{1,t}-\mu
_{\theta _{1}})(\theta _{2,t}-\mu _{i})]/\sum_{t=1}^{n}\pi _{i,t}^{\ast },
\end{equation*}%
\textit{where }$\pi _{i,t}^{\ast }=\Pr(K_{t}=i|\theta _{2,t})$ is\textit{\ }%
the probability of $\theta _{2,t}$ coming from the $i$-th component\textit{.}

\section{A clustering algorithm for fast estimation of mixtures 
of normals in adaptive IMH\label{s:clustering}}

Finite mixtures of normals are an attractive option to construct 
the proposal density because they can approximate any continuous 
density arbitrarily well and are fast to sample from and 
evaluate. See \cite{mclachlan00} 
for an extensive treatment of finite mixture models.

However, estimating mixtures of normals is already a difficult 
problem when an independent and identically distributed sample  
from the target is given and estimation needs to be performed only once: the likelihood goes to infinity
whenever a component has zero variance (an even more concrete possibility
if, as unavoidable in IMH, some observations appear more than once), and its
maximization, whether by the EM algorithm or directly, is plagued by local
modes. Although several authors have made substantial advances in dealing
with these problems \citep*[e.g.][]{figue02,ueda00, verbeek03}, in our experience 
the EM algorithm does not seem to be
sufficiently reliable when the sample is small and contains a 
non-trivial share of rejected draws. The inevitable short runs 
of rejections give rise to small clusters with zero covariance 
matrix at which the EM algorithm
frequently gets stuck. Moreover, EM-based algorithm are computationally
expensive and slow to converge, which makes them less attractive when the
proposal is to be updated frequently.

We have therefore limited our attention to algorithms that 
estimate mixtures
of normals quickly and without explicitly computing the 
covariance matrix of
each component (for robustness). Within this class, the 
\textit{k-means} is the most popular algorithm. We employ the \textit{k-harmonic means}, an
extension of the k-means algorithm that allows for soft 
membership. 
Degeneracies can be easily prevented, so the algorithm is 
remarkably robust
even in the presence of a long series of rejections. The number of 
clusters is
chosen with the BIC criterion. The increase in speed and 
reliability is paid
for with a decreased fit to the target, as the family of k-means algorithms
performs best when an optimal fit requires all components of the mixture to
have the same probability and covariance matrix \citep[see][for 
a discussion]{bradly98}.  \cite{hamerly02} 
show that the performance of k-harmonic means deteriorates less 
rapidly than alternatives of similar
computational cost with departures from these ideal conditions. An outline of 
the k-harmonic means algorithms is given in Appendix~1. 

Although the k-harmonic means algorithm is less sensitive to 
initialization
than either k-means or EM \citep{hamerly02}, 
in an unsupervised environment it is important to have good 
starting values. We have found the
algorithm of \cite{bradly98} 
to perform very well and at a low computational cost.

If the proposal distribution is normal then it is 
computationally inexpensive to
update it at every iteration. It is tempting to update a mixture 
of normals 
proposal with an on-line estimation procedure such as the on-line EM
algorithm proposed in \cite{andrieu06}. 
The advantage 
of on-line estimation is that the parameters of the mixture are updated recursively, so
the proposal itself is updated at each iteration at a very small
computational cost. However, on-line estimation of the mixture parameters in
AIMH has a number of serious shortcomings. The estimates are 
inefficient
compared to batch estimators because each data point is used 
only once,
which corresponds to requiring a batch optimization to converge in one step.
The loss of efficiency is more severe in small samples, that is, in the
early phases of the chain. Direct estimation of the mixture component
covariance matrices often leads to numerical problems in the early phases of
the chain given that rejections in MH produce degenerate clusters. Finally,
a limitation of on-line estimators is related to the fact that 
they are a form of stochastic gradient descent (see 
\cite{spall03} for an introduction). 
When the function to be maximized is multimodal (as typically the case with
mixtures) on-line estimates are in general sensitive to the order of the
draws, with initial draws having heavier influence than later draws in
determining the mode at which estimates settle down. We have verified
empirically that the quality of solutions given by several on-line
algorithms deteriorates rapidly if the initial observations are not
representative of the entire target distribution. This makes on-line
algorithms unsuitable for use in the early, exploratory phases of the chain,
though they can work well if the initial proposal distribution already
provides a reasonably good approximation of the target and the 
acceptance rates are sufficiently high. 

Since we are opting for batch estimators, it is too costly to 
update the proposal at each iteration. We update it at 
predetermined numbers of
iterations, more frequently in the earlier stages of the chain. 
Implementation details for the empirical examples are given in 
section~\ref{s:implementation}. 

We make two further comments on \cite{andrieu06}. First, they propose to keep the 
number of components in the mixture constant, whereas we have found it 
advantageous to select the number adaptively 
as outlined in appendix~1. Second, they outline 
a proposed approach to using mixtures as proposal densities, but do not report on 
the empirical performance of their proposal.

\section{Discussion} \label{s:discussion}
In order to understand the strengths and limitations of our sampler, we have
found it useful to consider two desirable qualities of an adaptive IMH
scheme. First, given a sufficiently large sample drawn from the target, we
wish to construct a proposal density which fits the target as well as
possible. This is an approximating ability: we want to draw an accurate
`map' of the areas that we have already explored. Second, we wish the
sampler to perform as well as possible when the initial proposal fails to
cover part of the support of the target distribution. This is an exploring
ability: when we propose in a region where our map is poor, we want to
explore that region and quickly update our map.

For example, using a normal proposal when the target is highly non-normal
results in little approximating ability. Updating the proposal only once or
very rarely results in little exploring ability, since the proposal reacts
slowly or not at all to the information that it is fitting poorly at a given
point.

Our sampler has several characteristics designed to enhance its exploring
ability. Frequent updating, particularly at early iterations, and updating
following a sequence of low MH acceptance probabilities
quicken the pace at which the proposal adapts to the information that
it is not fitting well in a certain area. Long tails are useful not only to
satisfy \eqref{e:domin} and \eqref{e:dimn adap},
but also to improve the chances of venturing into unexplored parts of the state 
space.
Finally, mixtures are ideally suited for this exploration because a new
component can be centered on a value causing a sequence of rejections. The
long runs of rejections that can plague standard IMH are therefore much less
likely using our AIMH sampling scheme because the proposal distribution is
updated frequently and will accommodate the cluster of rejections by
changing the mixture parameters or adding a new component. If our sampler
makes a move in a region where the proposal fits poorly, it should therefore
be able to explore it. Of course as the parameter dimension increases, if
the initial proposal fails to cover a region we may never explore that
region simply because the probability of making a proposal there is too
small.

The next example shows that in low dimensions we can often get away with a
very poor initial proposal distribution. 

\begin{figure}[tbp]
\centering
\includegraphics[angle=0,height = 2.5in, width=1.0\textwidth]{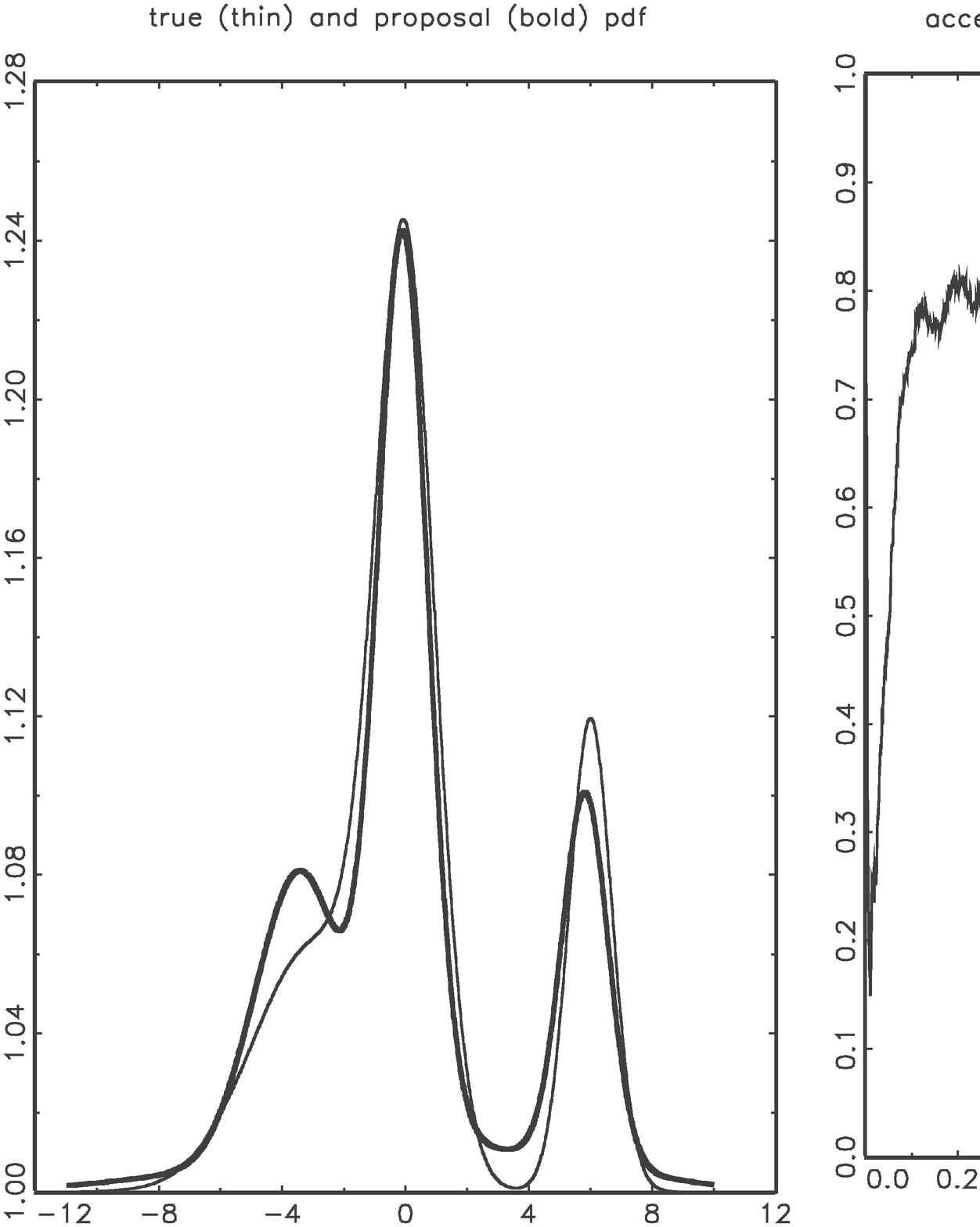}
\caption{Example~1. Left panel: Proposal
distribution (bold)  after 15 000 iterations, initialized with a normal $\protect%
\phi (z;-5,4)$. The target density (thin)  is a univariate mixture $0.5\protect\phi %
(z;0,1)+0.3\protect\phi (z;-3,4)+0.2\protect\phi (z;6,0.5).$ Right panel:
Recursive updates of the acceptance rate in the last 500 iterations.}
\label{Dim1}
\end{figure}

\noindent
{\sf Example~1} Suppose that the target is the
univariate mixture 
\begin{equation*}
\pi (z)=0.5\phi (z;0,1)+0.3\phi (z;-3,4)+0.2\phi (z;6,0.5),
\end{equation*}%
and the initial proposal is $\phi (z;-5,4).$ This proposal has very high
importance weights $\pi /g$ in a large part of the support of $z$, but we
still quickly converge to a good approximation of the target. Figure~\ref{Dim1} shows that 
the acceptance
rates increase fast initially and then stabilize as the proposal
distribution also settles down. 

The next example shows that as the dimension increases a good initial proposal distribution becomes
more valuable. 

\begin{figure}[tbp]
\centering
\includegraphics[angle=0,height = 4in, width=1.0\textwidth]{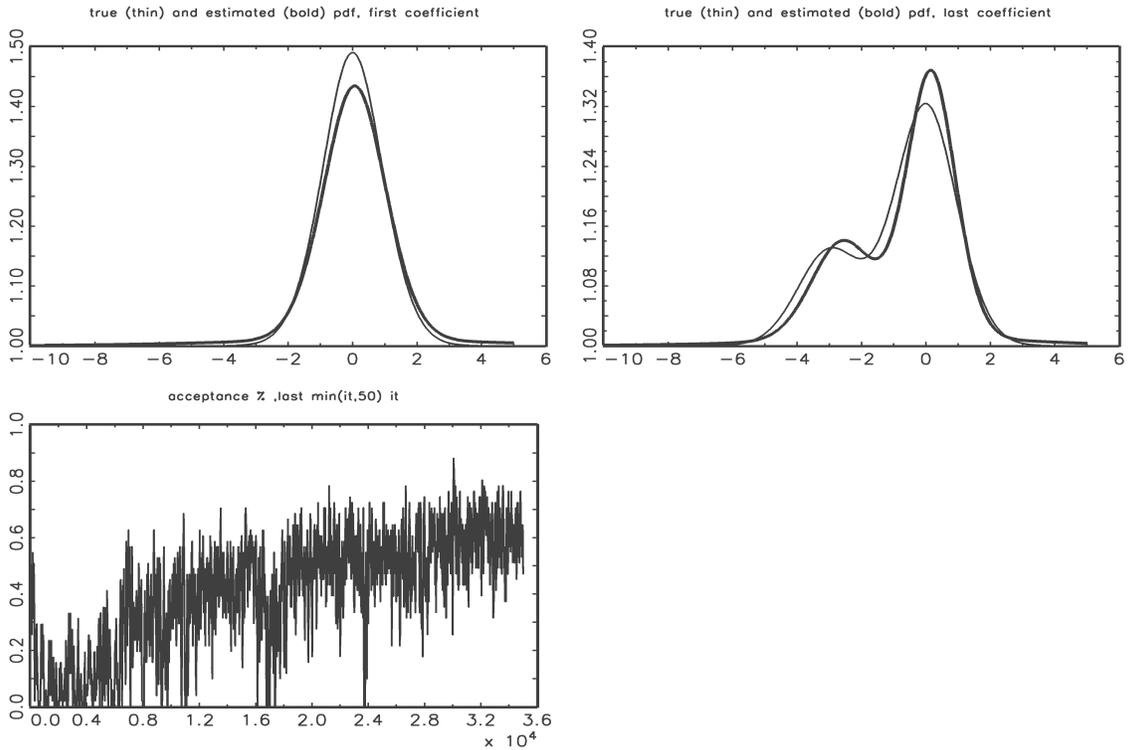}
\caption{Example~2. Proposal distribution after 35
000 iterations. The target distribution  is a 15-dimensional mixture. The
graph plots the true marginal distributions (thin)  for the first and last variable
together with the corresponding marginal distributions (bold)  implied by the
mixture of normals estimated on the full history of the draws, and with
recursive updates of the acceptance rate in the last 500 iterations.}
\label{Dim15}
\end{figure}

\noindent
{\sf Example~2} Consider the fifteen dimensional target
distribution which is the  mixture of two normals%
\begin{equation*}
\pi (z)=0.7\phi (z;0,I)+0.3\phi (z;\mu _{2},2I),
\end{equation*}%
where $\phi (z;\mu ,\Sigma )$ is a multivariate normal density with mean $%
\mu $ and covariance matrix $\Sigma $ evaluated at $z.$ The vector $\mu _{2}$
has elements $\mu _{2,i}=0$ for $i=1,...,14,$ and $\mu _{2,15}=-3.$ The
first fourteen marginals are therefore symmetric but slightly leptokurtic,
whereas the fifteenth is also skewed. The proposal distribution is
initialized by fattening and stretching the tails of the Laplace
approximation, that is, a normal distribution centered at the mode and with
covariance set to minus the inverse of the Hessian of the log-likelihood at
the mode. The Laplace approximation is nearly equal to $\phi (z;0,I)$, so we
have $g_{0}(z)\simeq 0.6\phi (z;0,I)+0.4\phi (z;0,16I)$. Figure~\ref{Dim15} 
shows that the acceptance rates at the initial proposal are not high, but sufficient to
start the learning process. The AIMH learns the
marginal distribution of the non-normal variable rather well and, since most
variables are normal, at very low computational cost since we only estimate
the mixture parameters on variables that appear skewed. In contrast, an
initial proposal such as $\phi (z;m,4I),$ where $m=-5$ generates such low acceptance rates for this fifteen dimensional
distribution that the learning process cannot get successfully started. 

\section{Applications\label{s:applications}}
State space models and nonparametric models are ideal initial  applications for
AIMH schemes. Although they can have a large number of parameters, it is
often the case that, conditional on a small subset, most parameters can be
integrated out or have known analytical form. Therefore it is often possible
to draw all parameters in one or two blocks. Exploiting these features, it
is also often inexpensive to find the posterior mode, possibly for a
simplified version of the model, and therefore obtain a reasonable
initialization of the proposal distribution. Finally, the standard approach
based on Gibbs and Metropolis-within-Gibbs can be very inefficient,
particularly for state space models \citep*[see][]{fruwirth04}.

For each of our applications we checked the results of the adaptive sampling scheme by re-running 
the sampler at a number of different starting points using a fixed proposal 
based on 
the last mixture in the \strict{} adaptation phase. In all cases we got very 
similar results to those obtained using \strict{} adaptation. 

For our examples we define the inefficiency of a sampling scheme as  
the factor by which the number of iterates would need to increase to 
give the same precision (standard error) as a sampler generating 
independent draws. For two sampling schemes A and B say, we define the 
inefficiency of scheme B relative to A as the factor by which it is necessary 
to increase the running time of B in order for it to 
obtain the same accuracy as A. It is computed as  the inefficiency factor of B times its run time per iteration divided by the inefficiency factor of A times its run time per iteration. 

In the examples below we compare the performance of the AIMH sampler to the 
following version of the \cite{haario01} adaptive random walk Metropolis sampler
proposed on page~3 of \cite{roberts06_applied}. Specifically, let $\theta$ be the parameters in the model, $\hat \theta$ the posterior mode and $V$ the variance covariance matrix of the Laplace approximation to the posterior. Then at iteration $j$ the proposal distribution is given by
\begin{align*}
Q_j(\theta^c,\cdot) & = N(\theta^c, (0.1)^2 V/d)  \quad \text{ if} \quad 
&  j <  5d, \\
Q_j (\theta^c, \cdot ) & = (1-\beta) N(\theta^c, (2.38)^2 \Sigma_j/d) + \beta 
N(\theta^c, (0.1)^2 I_d/d) \quad \text{if} \quad & j \geq  5d, 
\end{align*}
where $N(\theta, V)$ is the normal density with mean $\theta$ and covariance matrix $V$, $\theta^c$ is the current value of $\theta$, $d$ is the dimension of $\theta$, $\beta = 0.05$ and $\Sigma_j$ is the current empirical estimate of the covariance matrix of the target distribution based on the iterates thus far. In all cases we initialized this sampler at the posterior mode.  



\subsection{Time-varying parameter autoregressive models}

Consider the following time-varying parameter first order autoregressive
(AR(1)) process (the extension to a more general autoregressive process is
straightforward): 
\begin{equation} \label{Infl} 
y_{t} =c_{t}+\rho _{t}y_{t-1}+\sigma _{\epsilon }\epsilon _{t}
\, , \text{  }  
c_{t} =c_{t-1}+\lambda _{0}\sigma _{\epsilon }u_{t} \text{   and   }  
\rho _{t} =\rho _{t-1}+\lambda _{1}v_{t}, 
\end{equation}%
where $\epsilon _{t},u_{t},v_{t}$ are all $nid(0,1).$ The model has three
parameters $(\sigma _{\epsilon }^{2},\lambda _{0}^{2},\lambda _{1}^{2}),$
while $c_{0}$ and $\rho _{0}$ can be treated either as parameters or (our
choice) as states. Given conjugate priors (inverse gamma for the parameters,
and normal for $c_{0}$ and $\rho _{0}$), Gibbs sampling is straightforward
\citep{carter94}. 
\cite{fruwirth04} 
reports that based on the autocorrelations of the iterates, 
Gibbs sampling can be very inefficient for these models. 
In the following application we also find that the Gibbs draws are
highly autocorrelated and, by comparing posterior statistics from Gibbs
sampling and from our AIMH, we also find that the autocorrelations do not reveal
the full extent of the problem.

\subsubsection{Application: US CPI inflation} 
\label{sss:inflation} 

We apply the model to quarterly U.S. CPI inflation for the period 1960-2005
(184 observations).\footnote{%
Annualized quarterly CPI inflation, defined as $400(P_{t}/P_{t-1}-1),$ where 
$P_{t}$ is aggregated from monthly data (averages) on Consumer Price Index
For All Urban Consumers: All Items, seasonally adjusted, Series ID CPIAUCSL,
Source: U.S. Department of Labor: Bureau of Labor Statistics.} We use rather
dispersed inverse gamma priors for $\sigma _{\epsilon 
}^{2},\lambda
_{0}^{2},\lambda _{1}^{2}$ with a common shape parameter of 1. 
The scale
parameters are defined by setting the modes of the priors close 
to maximum
likelihood estimates: $\sigma _{OLS}^{2}$ for $\sigma _{\epsilon 
}^{2}$
(where $\sigma _{OLS}^{2}$ is the residual variance from an 
AR(1) model
estimated by OLS), at $0.01\sigma _{OLS}^{2}$ for $\lambda 
_{0}^{2}$ and at $%
0.001^{2}$ for $\lambda _{1}^{2}.$ The modes of $\lambda _{0}^{2}$ and $%
\lambda _{1}^{2}$ are centered at the maximum likelihood estimates to ensure
that the bimodality in the posterior distribution of the log of $\lambda
_{1}^{2}$ documented in Figure~\ref{InflMH} is not induced by the prior.

For given parameters, the likelihood is easily computed via the Kalman
filter. It is therefore simple to find the posterior mode, at which the
chain is initialized. 
Posterior mode values suggest that time variation is
nearly absent.

\begin{figure}[tbp]
\centering
\includegraphics[angle=0,height = 3.5in, width=1.0\textwidth]{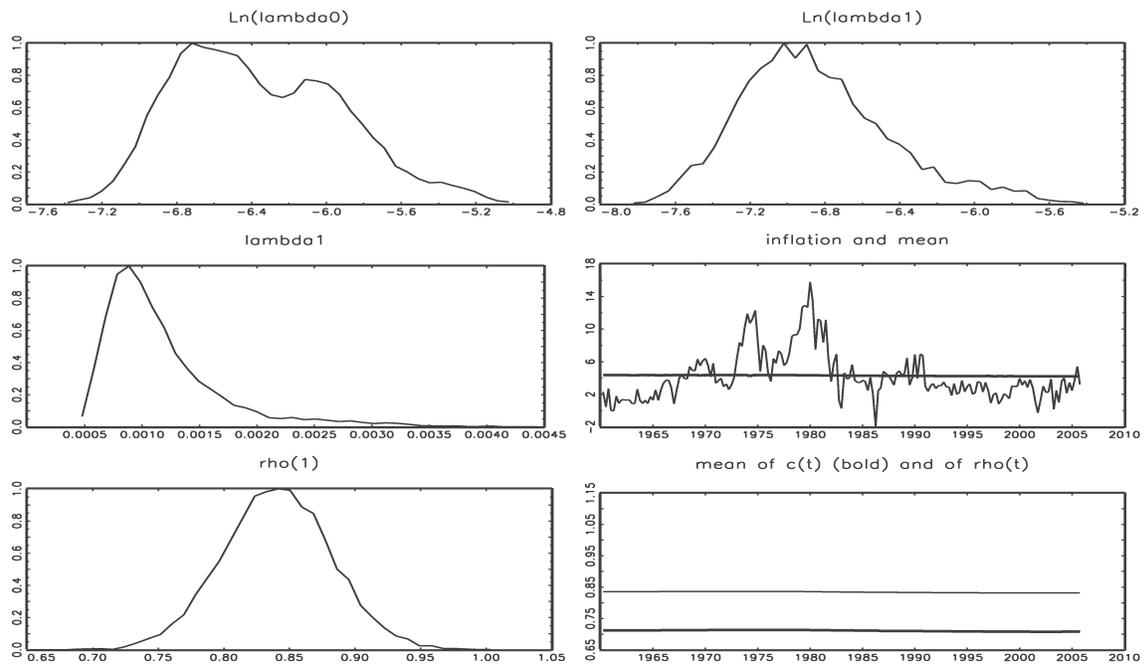}
\caption{Inference for a time varying
parameter AR(1)\ model for US inflation by Gibbs sampling. (a) marginal
distribution of $\ln (\protect\lambda _{0})$ (b) marginal distribution of $%
\ln (\protect\lambda _{1})$ (c) marginal distribution $\protect\lambda _{1}$
(d) inflation plot and mean, estimated as $E[(c_{t}/1-\protect\rho _{t})|y]$
(e) marginal distribution of $\protect\rho _{0}|y$ (f) $E(c_{t}|y)$ (bold
line) and $E(\protect\rho _{t}|y).$}
\label{InflGibbs}
\end{figure}

\begin{figure}[tbp]
\centering
\includegraphics[angle=0,height = 3in, width=1.0\textwidth]{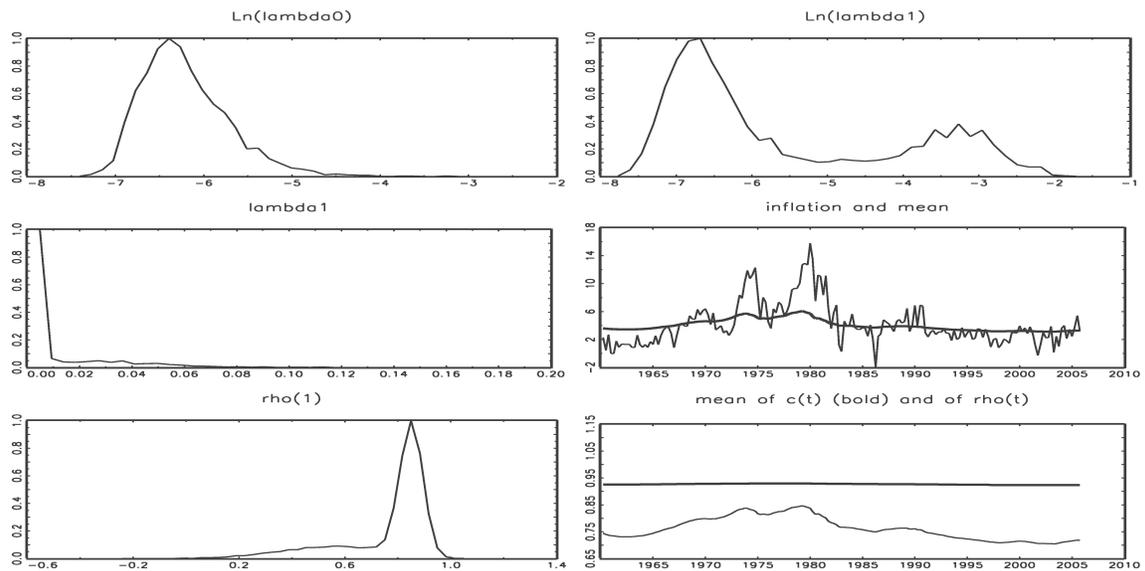}
\caption{Inference for the model of
figure \protect\ref{InflGibbs} by adaptive IMH. The interpretation of the
panels is the same as in figure \protect\ref{InflGibbs}.}
\label{InflMH}
\end{figure}

Starting with Gibbs sampling, we draw 40 000 times after a   
burn-in of 5000. 
The recursive parameter means seem to settle down (not reported) and the
posterior distributions are in line with a normal approximation taken at the
mode, suggesting a persistent AR(1) with little sign of parameter variation
(see Figure \ref{InflGibbs}). It may therefore seem reasonable to assume
that the chain has produced a sample representative
of the entire posterior.

However, the AIMH scheme tells a different story. The proposal is
initialized at a mixture of two normals 
$g_{0}(z)=0.5\phi (z;\widehat{\mu },\widehat{\Sigma })+0.5\phi (z;\widehat{%
\mu },16\widehat{\Sigma })$,
where $\widehat{\mu }$ is the posterior mode and $-\widehat{\Sigma }$ is the
inverse of the Hessian of the log-posterior evaluated at $\widehat{\mu }$.
The AIMH soon discovers that the posterior distribution of $\log (\lambda
_{1}^{2}),$ not to mention $\lambda _{1}^{2}$, is highly non-normal (see
Figure~\ref{InflMH}), with substantial probability mass around a second mode
corresponding to non-trivial amounts of time variation in $\rho _{t}$ and a
lower $\rho _{1}.$ 

We also ran the adaptive random walk Metropolis sampler outlined at the start 
of the section. The sampler settles down to an acceptance rate of 20\% and 
obtains the correct posterior distribution, and in particular finds both 
modes. Table~1 gives the inefficiency factors for all three samplers as well 
as the inefficiency factors of the Gibbs and ARWM relative to AIMH. The table 
shows that the AIMH sampler is appreciably more efficient than the other 
two samplers. 

\subsection{Additive semiparametric Gaussian models\label{Sec:
Semiparametric}} \label{ss:semipar} 

In this example we consider the additive semiparametric 
regression model with Gaussian
errors, with some of the covariates entering linearly and the 
others entering more flexibly 
\begin{equation}
y_{i}=\sum_{j=1}^{m}\gamma _{j}z_{ji}+\sum_{h=1}^{H}f_{h}(x_{h,i})+\sigma
_{\epsilon }\epsilon _{i} \, ;   \label{eq: Nonparam1}
\end{equation}%
the $\epsilon _{i}$ are $nid(0,1)$ and $z$ is a vector of regressors that
enter linearly. The $x_{h},$ $h=1,...,H$ are covariates that enter more
flexibly by using the quadratic polynomial spline functions 
\begin{equation} \label{eq: Nonparam2}
f_{h}(x_{h,i}) =\beta _{0,h}x_{h,i}+\sum_{j=1}^{J}\beta _{h,j}(x_{h,i}-%
\tilde{x}_{h,j})_{+}^{2} = \beta _{0,h}x_{i}+g_{h}(x_{h,i}), 
\end{equation}
where $x_{+}=x$ if $x>0$ and 0 otherwise and $\{\widetilde{x}_{h,1},...,%
\widetilde{x}_{h,k}\}$ are points (or `knots') on the abscissae of $x_{h}$
such that $\min (x_{h})=\widetilde{x}_{h,1}<...<\widetilde{x}_{h,J}<\max
(x_{h})$. In this paper we choose 30 knots so that each interval contains
the same number of observed values of $x_{h}.$ For a discussion of quadratic
spline bases and other related bases see chapter~3 of \cite{ruppert03}. 
We assume that a global intercept term is included in $z$ in
(\ref{eq: Nonparam1}) and for simplicity we include the parameters $\beta
_{h,0},$ $h=1,\dots ,H$ in the vector $\gamma $ and $x_{h},$ $h=1,\dots ,H$
as part of the vector $z$. This transforms the nonparametric model into an
highly parametrized linear model

\begin{equation}
y=\widetilde{Z}\widetilde{\gamma }+\epsilon .  \label{eq: NonParam}
\end{equation}%
The prior for the linear parameters $\gamma $ is normal with a diagonal
covariance matrix $\gamma \sim N(0,v_{\gamma }^{2}I)$,
where\ $v_{\gamma }$ can be set to a large number. It is also convenient to
assume a normal prior for the nonparametric part, with all parameters
independent and 
$\beta _{h,j}\sim N(0,\tau _{h}^{2}),\ , j=1,...,J,\,  h=1,...,H.$
However, with this prior there is a high risk of over-fitting if we
simply set $\tau _{h}^{2}$ to a large number. The variance $\tau _{h}^{2}$
is often chosen by cross-validation, but in a fully Bayesian setting we can
treat $\tau _{h}^{2}$ as a parameter. To illustrate the advantage of AIMH in
working with different priors, we experiment with two options for the prior $%
\tau _{h}^{2}$. The first prior is log-normal 
and rather dispersed:~$\ln (\tau _{h}^{2})\sim N(0,5^{2})$,
the second is inverse gamma with shape parameter 1 and scale parameter
implied by setting the mode at $0.1^{2}$. The prior for $\sigma _{\epsilon
}^{2}$ is inverse gamma with shape parameter one and scale parameter implied
by setting the prior mode at the OLS residual variance estimated on (\ref%
{eq: NonParam}). The prior for $\widetilde{\gamma }=(\gamma ,\beta
_{1},...,\beta _{H})$ is therefore jointly normal conditional on $\tau
^{2}=\{\tau _{1}^{2},..,\tau _{H}^{2}\}$, %
$\widetilde{\gamma }|\tau \sim N(\mathbf{0},V_{\widetilde{\gamma }}(\tau ))$, 
where $V_{\widetilde{\gamma }}(\tau ) = \text{diag} (v_\gamma^2 I, \tau_1^2, I, \dots, \tau_H^2I) $ 
is a block diagonal matrix. 
One way to estimate the posterior density of the semiparametric model is to
use Gibbs or Metropolis-within-Gibbs sampling as proposed by \cite{wong96}. 
In this approach the parameters $\widetilde{\gamma }%
=\{\gamma ,\beta _{1},...,\beta _{H}\}$\ are conjugate given $\theta
=\{\sigma _{\epsilon }^{2},\tau _{1}^{2},...,\tau _{H}^{2}\},$\ and $\sigma
_{\epsilon }^{2}$\ is conjugate given $\widetilde{\gamma }.$\ Each variance $%
\tau _{h}^{2}$\ can be updated with a Gibbs step for the inverse gamma
prior, or with a Metropolis-Hastings step for the log-normal prior. In this
second case, we use a Laplace approximation of $p(\ln (\tau _{h}^{2})|\beta
_{h})$, which is very fast to compute using analytical derivatives. However,
the correlation between $\tau _{h}^{2}$ and $\{\beta _{h,1},..,\beta
_{h,J}\} $ could be quite high using either prior for $\tau _{h}^{2}$. In
addition, using a log normal prior for $\tau _{h}^{2}$ leads to high
rejection rates in the Metropolis-Hastings step when generating the $\tau
_{h}^{2}$. Both problems are elegantly solved by integrating out $\widetilde{\gamma }$ and generating 
$\theta$ as a block using an efficient AIMH sampler.

The next example shows how to update all parameters in one block with an efficient AIMH
sampler. We first note that, conditional on $\theta $, $\widetilde{\gamma }$
can be integrated out, making it possible to compute $p(\theta |y)\propto
p(y|\theta )p(\theta ),$ where $y|\theta \sim N(\mathbf{0},\sigma _{\epsilon
}^{2}I+\widetilde{Z}V_{\widetilde{\gamma }}(\tau )\widetilde{Z}^{\prime })$. 

\subsubsection{\protect\bigskip Application: Boston housing data}

We use the Gaussian semiparametric model to study the Boston housing data introduced by \cite{harrison78} and 
analyzed semiparametrically by \cite{smith96}. 
\footnote{The dataset is
available at www.cs.utoronto.ca/\symbol{126}delve/data/boston.} There are
506 observations. The dependent variable is the log of $MV$, the median
value of owner-occupied homes. We use all 13 available covariates (see Smith
and Kohn or the web-site for a full description) in the linear part and the
following six in the nonparametric part (Smith and Kohn use only the first
five): $X_{5}=NOX$, nitrogen oxide concentration,
$X_{6}=RM,$ average number of rooms,
$X_{8}=DIS,$ logarithm of the distance from five employment centers,
$X_{10}=TAX,$ property tax rate,
$X_{13}=STAT,$ proportion of the population that is lower status,
$X_{1}=CRIM,$ per capita crime rate by town.

\begin{figure}[tbp]
\centering
\includegraphics[angle=0, width=1.0\textwidth]{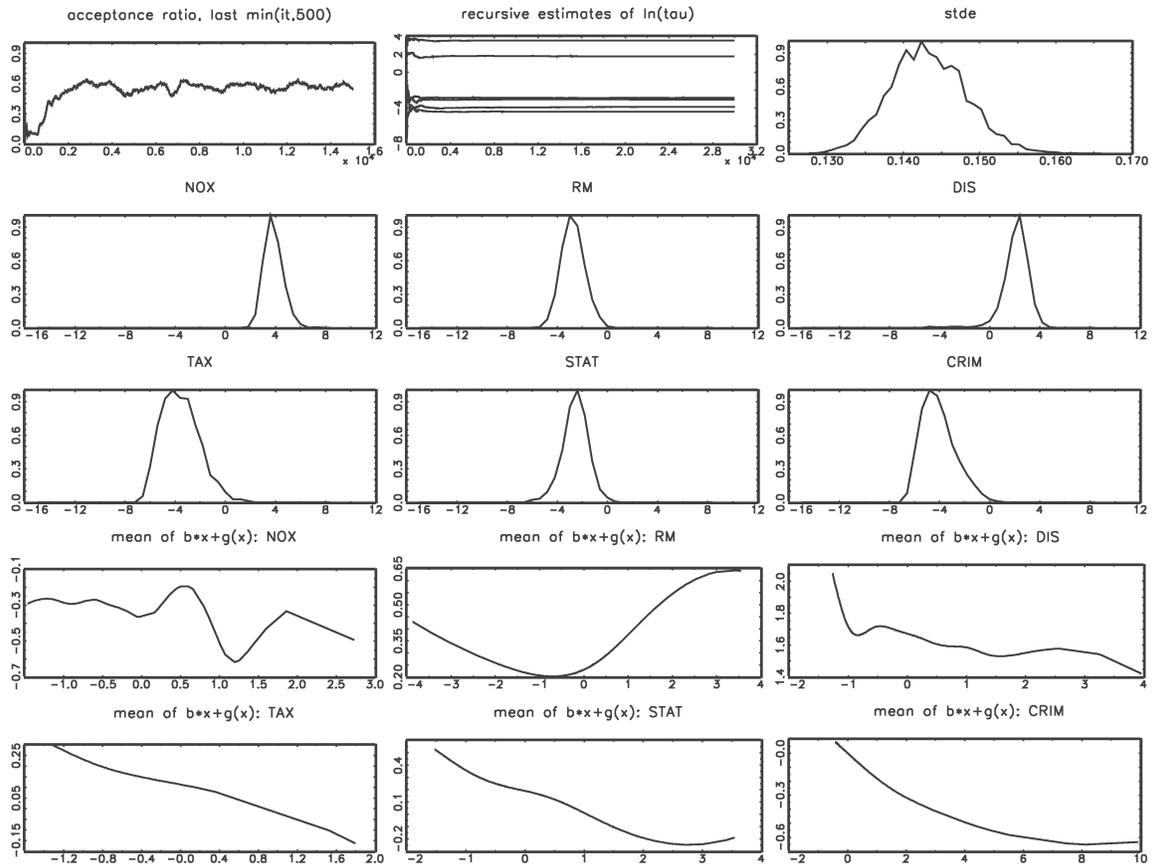}
\caption{Inference for semiparametric
model of housing prices by adaptive IMH. First row: recursive acceptance
rate for the last min(it,500) iterations, recursive means of $\ln (\protect%
\tau _{i}),$ marginal of $\protect\sigma _{\protect\epsilon }.$ Second and
third rows: marginals of $\ln (\protect\tau _{i}).$ Fourth and fifth rows:
means of $\protect\beta _{i}x+g_{i}(x).$}
\label{BostonMH}
\end{figure}

The proposal distribution for the seven parameters $\theta =\{\ln (\sigma
_{\epsilon }^{2}),\ln (\tau _{5}^{2}),...,\allowbreak \ln (\tau _{1}^{2})\}$
is initialized by fattening the tails of the Laplace approximation. To find
the Laplace approximation, we simply apply Newton-Raphson optimization (with
numerical derivatives) to $\ln p(y|\theta )+\ln p(\theta ),$ which involves
no extra coding effort since both densities are needed to compute the MH
acceptance ratio. Figure~\ref{BostonMH} provides results for the
case of a log-normal prior on $\tau _{h}^{2},$ $h=1,...,H$ and shows that the acceptance
rate quickly improves and stabilizes at around 60\% when all seven parameters are updated jointly. 
Most parameters are
approximately lognormally distributed, except those connected to the
variables $TAX$ and $CRIM$, which benefit from the added flexibility of
mixtures.\ The correlation matrix of the smoothing parameters $\{\ln (\tau
_{5}^{2}),...,\ln (\tau _{1}^{2})\}$ is nearly diagonal. This suggests that
the AIMH could handle large numbers of smoothing parameters efficiently by
updating them in blocks (with a different proposal density estimated
adaptively on each block), since the blocks would be nearly independent of
each other.

Table~\ref{t:table1} reports the inefficiency factors for both the 
Gibbs sampler and the AIMH sampler for both inverse gamma and 
log normal priors, as well as the inefficiency of the Gibbs sampler 
relative to the AIMH sampler. 
The table shows that in terms of relative efficiency (defined at the beginning 
of section~\ref{s:applications}), 
the AIMH is about 40\% more efficient than  
the Gibbs sampler when
both samplers use the inverse gamma prior on $\tau _{h}^{2}$, 
and nearly seven times more efficient when both samplers use the 
log-normal prior. Reported results
are for the average inefficiency factors (over both $h$ and $i$) of $%
f_{h}(x_{h,i})$. Looking at the autocorrelation of the log-parameters gives
similar inefficiency ratios.

We also applied the adaptive random walk Metropolis sampler to this data set, 
but could not make it work well. With the sampler initialized at the posterior 
mode, the acceptance rate started at over 50\%, but 
within a few hundred iterations fell to below 1\% and stayed there 
indefinitely. We do not report any inefficiency factors for 
this sampler because we do not believe that inference is reliable with such a 
low acceptance rate. We conjecture that the poor behavior of the ARWM sampler 
in this example compared to the other two examples is because
this example has 7 
parameters whereas the other two have 3 and 2 parameters. In addition, the 
second derivatives of the log posterior in this example are far from constant, 
so a unique covariance matrix may do very poorly. By contrast, a mixture of 
normals allows for local correlations between the parameter and therefore may 
be less affected. 

\begin{table}[htb] 
\centering%
\begin{tabular}{|l|l|l|l|l|l|l|}
\hline
\textbf{Boston} & mean $f_{h}(x_{h,i})$ &  & \textbf{Inflation} & $\log
(\sigma _{\epsilon }^{2})$ & $\log $($\lambda _{0}^{2})$ & $\log $($\lambda
_{1}^{2})$ \\ \hline
AIMH, IG & 2.6   &  & AIMH & 6.7 & 2.8   & 6.1   \\ 
Gibbs, IG & 6.3 (1.4) &  & Gibbs & 9.4 (1.3) & 113.3 (37.4)& 156.4 (23.7)\\ 
AIMH, LN & 1.6 (6.8)  &  & ARWM & 21.5 (3.1)  & 23.5 (8.3) &  23.6 (3.8) \\ 
M-Gibbs, LN & 18.4 &  &  &  &  &  \\ \hline
\end{tabular}%
\caption{Inefficiency factors for the semiparametric (Boston) and state space 
(inflation) models, together with the inefficiencies of the Gibbs sampler and the ARWM relative to the AIMH sampler in brackets.   
AIMH: adaptive independent Metropolis-Hastings; 
M-Gibbs: Metropolis-within-Gibbs; and ARWM: adaptive random walk Metropolis. 
IG and LN: inverse gamma and log-normal priors for the Boston data.
\label{t:table1}}
\end{table}

\subsection{Stochastic volatility models}

The simplest stochastic volatility model can be written for mean corrected
data as%
\begin{equation} \label{eg: SV}
y_{t} =e^{0.5h_{t}}\epsilon _{t} \, , \text{   } 
h_{t} =\mu +\rho (h_{t-1}-\mu )+\sigma v_{t},  
\end{equation}%
where $\epsilon _{t}$ is $nid(0,1)$ and the model parameters are $\theta
=\{\mu ,\rho ,\sigma \}.$ We square and take logs of the observation
equation, and we approximate the distribution of $\ln (\epsilon _{t}^{2}),$
which is the log of a chi-squared 1, by a mixture of normals as in Kim et
al. (1998). This model has a conditionally Gaussian state space form%
\begin{equation}
\widetilde{y}_{t} = g(K_{t})+h_{t}+G(K_{t})u_{t}\, , \text{   } 
h_{t} = \mu +\rho (h_{t-1}-\mu )+\sigma v_{t}, 
\end{equation}%
where $\epsilon _{t}$ is $nid(0,1),$ $\widetilde{y}_{t}=$ $\ln (y_{t}^{2}),$
and $g(K_{t})$ and $G(K_{t})$ are known given $K_{t}$.

The indicators $K$ can be sampled in one block given $\theta $ and $h$ as in
Carter and Kohn (1994). The distribution of $\theta $ given $h$ is
conjugate, but \cite{kim98} show that $\theta $ and $h$ are highly
correlated and recommend drawing $\theta $ given $y$ and $K$ but integrating 
$h$ out. This is accomplished with a Metropolis-Hastings step, where $%
p(y|K,\theta )$ is computed via the Kalman filter. Since the posterior mode
is not readily available, \cite{kim98} 
use IMH, where the proposal
distribution is calibrated once from draws obtained with a less efficient
sampling scheme. This is less efficient than our scheme and requires coding
two different samplers. An alternative we now explore is to use AIMH from
the beginning of the chain.

\subsubsection{Application: USD-GBP daily returns}

\begin{figure}[tbp]
\centering
\includegraphics[angle=0, width=1.0\textwidth]{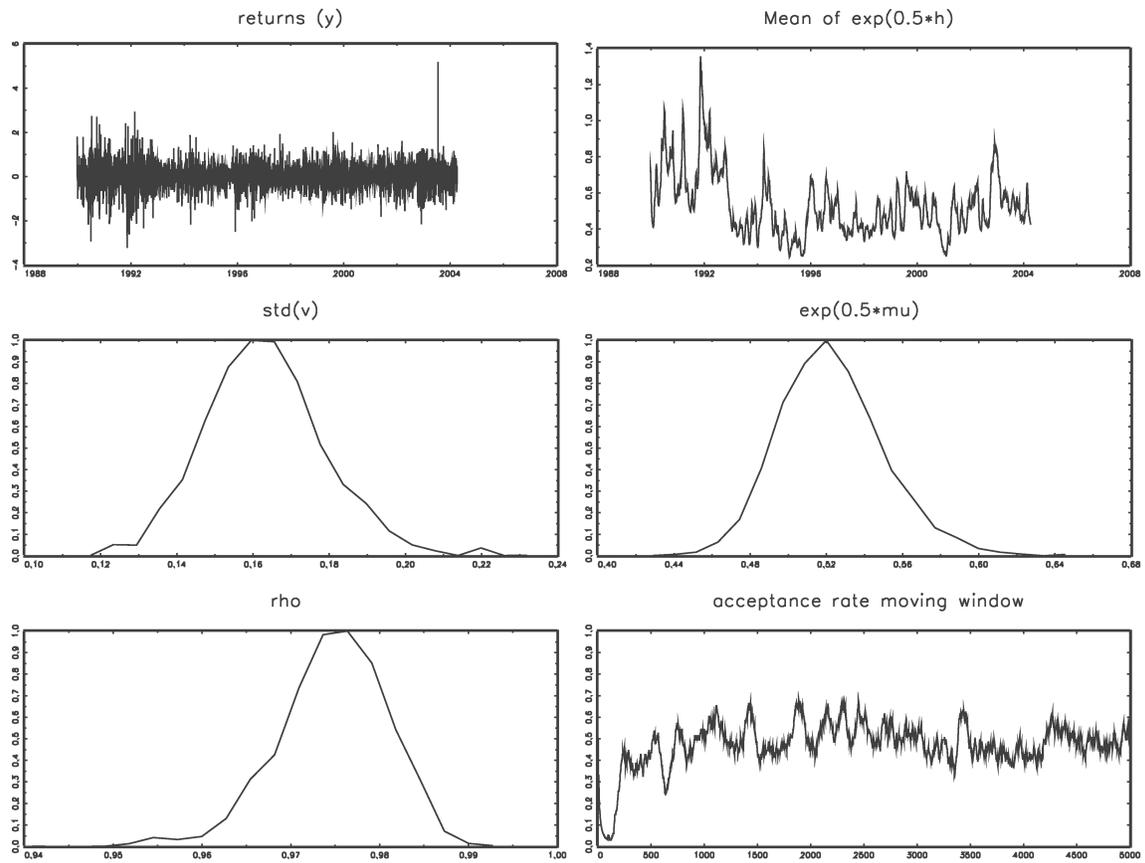}
\caption{Inference for the daily US-GBP
exchange rate by AIMH. (a) exchange rate returns (b) mean of $0.5\ln (h_{t})$
(c) marginal of $\protect\sigma _{v}$ (d)\ marginal of $0.5\exp (\protect\mu %
)$ (e) marginal of $\protect\rho $ (f) moving window of the acceptance rate
for the last min(it,500) iterations.
}
\label{sv}
\end{figure}

We analyze daily U.S. dollar - British pound returns (defined as the first
difference of the log of the nominal exchange rate) for the period January
1990 to March 2004. The parameter $\mu $ can be integrated out (see \cite{kim98}).  
To initialize the proposal distribution, we approximate the
distribution of $\ln (\epsilon _{t}^{2})$ as a normal with mean $-1.27$ and
variance $2.22^{2}$. This gives a standard Gaussian state space
model, for which the Laplace approximation is easily available. We also use
the mode to center the priors for $\rho $ and $\ln (\sigma ^{2})$, which are
normal and dispersed. The prior for $\rho $ is truncated at 1. With fattened
tails, the initial proposal gives an acceptance rate of around 10\%, and 
Figure~\ref{sv} shows that it
takes only a few hundred iterations for the acceptance rates to increase to
around 50\%.  This number is satisfactory given that
the proposal approximates $p(\theta |y)$ while the acceptance probability is
computed on $p(\theta |y,K)$. 

The adaptive random walk Metropolis sampler also quickly settled down to an acceptance rate of around 15\%, with the chain mixing well. The inefficiency factors for $\rho $ and $\sigma^2$ for both samplers are given in table~\ref{t:table2}, together with the inefficiency factors for the adaptive random walk Metropolis relative to AIMH. The table shows that AIMH compares favourably with ARWM.  

\begin{table}[htb] 
\centering%
\begin{tabular}{|l|l|l|}
\hline
\textbf{Sampling Scheme } & $\rho$ & $  \sigma^2$ \\ \hline 
AIMH  & 7.6   &  9.2 \\ 
ARWM & 21.1 (2.7)  & 26.5 (2.86)\\ \hline 
\end{tabular}%
\caption{Inefficiency factors for the stochastic volatility model for the adaptive independent Metropolis-Hastings (AIMH) sampler and the 
adaptive random walk Metropolis (ARWM) sampler, with the relative inefficiency of the AWRM in brackets. 
\label{t:table2}}
\end{table}

\section{Conclusion}
This paper shows that it is possible to build 
\aimh{} samplers that can do better than two-step adaptation because they adapt 
throughout the sampling period. The most interesting applications arise when 
current best practice is inefficient or cumbersome and, in our opinion, when 
adaptation starts early. 
Our article provides a fast and reliable algorithm which 
performs well in three interesting models and compares favorably on these 
examples with a standard Markov chain Monte Carlo sampler and an adaptive 
random walk Metropolis sampler. 

\section*{Acknowledgement} We would like to thank Luke Tierney, Christophe 
Andrieu and Antonietta Mira for helpful suggestions and 
questions that helped improve the accuracy and presentation of a 
previous version of the paper. Robert Kohn's research was partially supported 
by an ARC grant.

\section*{Appendix~1: k-harmonic means clustering} \label{s:k-harmonic}

We estimate the mixture of normal parameters using the k-harmonic means
clustering algorithm which can be described as follows. \citep[See][for a discussion]{hamerly02}. Let $p$ be the number of clusters.

\begin{enumerate}
\item Initialize the algorithm with $c_{1},...,c_{p},$ the component
centers. The starting values are chosen with the procedure of \cite{bradly98} . 
We depart slightly from Bradley and Fayyad in using the
harmonic k-means algorithm (rather than k-means) in the initialization
procedure.

\item For each data point $\theta _{t},$ compute a weight function $w(\theta _{t}) $ and a membership function $%
m(c_{i}|\theta _{t})$ for $t=1,...,p $ as 
\begin{equation*}
w(\theta _{t})=\frac{\sum_{i=1}^{p}||\theta _{t}-c_{i}||^{-p-2}}{%
(\sum_{i=1}^{p}||\theta _{t}-c_{i}||^{-p})^{2}} \quad \text{and} \quad 
m(c_{i}|\theta _{t})=\frac{||\theta _{t}-c_{i}||^{-p-2}}{\sum_{i=1}^{p}||%
\theta _{t}-c_{i}||^{-p-2}},
\end{equation*}%
where $||\theta _{t}-c_{i}||$ is the Euclidean or Mahalanobis distance.
Following \cite{bradly98}, 
we put a lower boundary $\epsilon $ on $%
||\theta _{t}-c_{i}||$ (to avoid degeneracies when trying $||c_{i}-c_{i}||$%
). The membership function softens the sharp membership of the k-means
algorithm, so one observation can belong to more than one cluster in
differing degrees. The weight function gives more weight to observations
that are currently covered poorly (i.e. that are far from the nearest
center).

\item Update each center $c_{i}$ 
\begin{equation*}
c_{i}=\frac{\sum_{t=1}^{n}m(c_{i}|\theta _{t})w(\theta _{t})\theta _{t}}{%
\sum_{t=1}^{n}m(c_{i}|\theta _{t})w(\theta _{t})}.
\end{equation*}

\item Repeat until convergence. This gives the cluster centers, which we
take as estimates of the component means. The other mixture parameters can
then be\ estimated for $i=1,...,k$ as%
\begin{equation*}
V_{i} =\frac{\sum_{t=1}^{n}m(c_{i}|\theta _{t})w(\theta _{t})(\theta
_{t}-c_{i})(\theta _{t}-c_{i})^{\prime }}{\sum_{t=1}^{n}m(c_{i}|\theta
_{t})w(\theta _{t})} \quad \text{and} \quad 
\pi _{i} \varpropto \sum_{t=1}^{n}m(c_{i}|\theta _{t})w(\theta _{t}).
\end{equation*}

\item The number of clusters is chosen with the BIC criterion given a
maximum number (5 in our examples).
\end{enumerate}

We notice that the covariance matrices $V_{i}$ are only estimated once,
after convergence. k-means type algorithms also differ from the EM algorithm
in that they do not evaluate the likelihood $p(\theta |c_{1},...,\pi
_{1,}V_{1},...)$. This sub-optimal use of information in fact turns out to
be a great advantage for our purposes. Fewer iterations than for EM are
needed for convergence, and each iteration is faster. Even more importantly,
the algorithm does not get stuck in the small degenerate clusters caused by
rejections in the sense that, unlike for the EM algorithm with freely
estimated covariances, these small clusters are not absorbing. If k-harmonic
means does find a degenerate cluster, this causes no trouble for
convergence, and after convergence we can use a predefined matrix in place
of any non-positive-definite covariance matrix (for example, if $V_{i}$ is
not positive definite we set it to $0.5^{2}Var(\theta )$). If desired, the
mixture parameters can be refined with a few steps of the EM algorithm. In
this case, we recommend not updating the the covariance matrices for the
reasons just discussed.

\section*{Appendix~2: Proofs} 

The one-step transition kernel for $Z_{n+1}$ in section~\ref{S: theory} 
is given by 
\begin{align} \label{e:transition}
T_n (z,d\zprime) & = \alpha_n(z,z^\prime) q_n(\zprime)\mu\{d\zprime\} + 
\delta_z(d\zprime) (1 - \nu_n(z))  
\end{align}
where $\delta_z(d\zprime) = 1$ if $z \in d\zprime$ and is 0 otherwise, and 
\begin{align} \label{e:def nu} 
\nu_n(z) = \int_{\calZ}\alpha_n(z,z^\prime) q_n(z^\prime)\mu\{ d\zprime\}.  
\end{align}
By the construction of the \MH{} transition kernel, 
\begin{align}\label{e:T invar} 
 \quad \int_{ \calZ} \pi(z) T_n(z,d\zprime)\mu\{dz\}  & = \pi(\zprime)
 \mu\{d\zprime\} \ . 
\end{align}

In this section $K$ is a generic constant, independent of $n,z$ and $\zprime$. It
is convenient to write $h_n(z; \lambda_n) $ 
as $h_n(z)$
Without loss of generality we assume throughout this section that
$\calZ$ is a discrete space. Exactly the same proof goes through for the
continuous case with summations replaced by integrals. We use the notation
$z_{s:t} $ to mean $\{z_s, \dots, z_t\}$ for $ s \leq t$, with a similar
interpretation for $Z_{s:t}$. 

To prove Theorem~1 
we first obtain the following two lemmas. 
\setcounter{theorem}{0}
\begin{lemma} \label{lm:lemma1}
Under the assumptions of Section~2, for any $n,k>0$ and $z,\zprime \in \cal Z$, \\
(a) $q_n(z)  \leq Kg_0(z)$. \\
(b) $\alpha_n(z,\zprime)q_n(\zprime)  \leq Kg_0(\zprime ) $\\
(c) There exists an $\eps_1$,  $0 <  \eps_1 < 1 $,  such that $\alpha_n(z,\zprime)q_n(\zprime) > \eps_1 \pi(\zprime) $ for all $z, \zprime \in \calZ$. \\
(d) $\nu_n(z ) > \eps_1$ for all $z \in \calZ$, where $\nu_n(z)$ is defined by \eqref{e:def nu}. \\
(e) For $k \geq 1  $, let 
$
\Delta_{n} (z, \zprime) = \alpha_n(z,\zprime) q_n(\zprime) - \alpha_{n+1}(z,\zprime) q_{n+1}(\zprime)$. Then, 
\begin{align} \label{e:dimn adap delta}
\mid \Delta_{n}(z,\zprime)  \mid \ \leq \ K \biggl( g_0(\zprime) + \frac{\pi(\zprime)}{\pi(z)} g_0(z) \biggr ) a_n. 
\end{align}
(f) \begin{align} \label{e:dimn adap nu} 
\mid \nu_n(z) - \nu_{n+1} z)\mid \leq K \biggl ( 1 + \frac{g_0(z}{\pi(z)} \biggr ) a_n 
\end{align}
\end{lemma} 

\begin{proof} [Proof] (a) $ q_n(z)/g_0(z) = \omega_1 + (1-\omega_1)\mixt_n(z)/g_0(z) $ and the result follows from \eqref{e:domin}. (b)~follows from (a) and $\alpha_n(z,\zprime) \leq 1$. To show (c), note that $q_n(z)/\pi(z) \geq \omega_1 g_0(z)/\pi(z) $. From~\eqref{e:domin}, there is an $\eps_1 $ such that $q_n(z)/\pi(z) > \eps_1$ for all $z \in \calZ$. It is now straightforward to show that $\alpha_n(z, \zprime) q_n(\zprime)/\pi(\zprime) > \eps_1$ for all $z, \zprime \in \calZ$.  (d) follows from 
\begin{align*}
\nu(z) & = \sum_{\zprime} \alpha_n(z, \zprime) q_n(\zprime) > \eps_1 \sum_{\zprime} \pi(\zprime) = \eps_1 
\end{align*}

To obtain (e), it is necessary to consider the following four cases. 

Case 1. $\alpha_n(z,z^\prime) = 1$ and  $\alpha_{n+1}(z,z^\prime) = 1$. Then, 
$| \Delta_{n} | = \mid q_n(\zprime) - q_{n+1} (\zprime) \mid \ \leq\ K  g_0(z^\prime) a_n $ by \eqref{e:dimn adap}.  

Case 2. $\alpha_n(z,\zprime) < 1$ and  $\alpha_{n+1}(z,\zprime) < 1$. 
\begin{align*}
\mid \Delta_n (z,z^\prime)  \mid & = \frac{\pi(\zprime)} {\pi(z)} \mid q_n(z) - q_{n+1} (z) \mid\  \leq \ K \frac{\pi(\zprime)} {\pi(z)} g_0(z) a_n . 
\end{align*}

Case 3.  $\alpha_n(z,z^\prime) = 1$ and  $\alpha_{n+1}(z,z^\prime) < 1$. In this case $\Delta_n (z,z^\prime) = q_n(z^\prime) - \pi(z^\prime) q_{n+1} (z) / \pi(z)$. If $\Delta_n (z,z^\prime) \geq 0 $, then 
\begin{align*} 
0 \leq\Delta_n (z,z^\prime) \leq \frac{\pi(z^\prime)} {\pi(z)} \biggl ( q_n(z) - q_{n+1} (z) \biggr ) \leq K g_0(z) a_nc_k \ . 
\end{align*} 
If $\Delta_n (z,z^\prime) < 0 $, then 
\begin{align*}
0 < -\Delta_n (z,z^\prime) = \frac{\pi(z^\prime)} {\pi(z)}q_{n+1}(z) - q_n(z^\prime) \leq
q_{n+1} (z^\prime) - q_n(z^\prime) 
\end{align*}
Thus, 
\begin{align*}
\mid \Delta_n (z,z^\prime) \mid \ \leq \ K \biggl ( g_0(\zprime)  + \frac{ \pi(\zprime) }{ \pi(z) } \biggr ) a_n . 
\end{align*} 

Case 4. $\alpha_n(z,z^\prime) < 1$ and  $\alpha_{n+1}(z,z^\prime) = 1$. This case is similar to case 3. 

To obtain (f), we note that 
\begin{align*}
\mid \nu _n(z) - \nu_{n+1} (z) \mid \ \leq\  \sum_{\zprime} | \Delta_{n} (z, \zprime) |  \ , 
\end{align*}
and the result follows from (e). 
\end{proof}

With $\eps_1$ as in Lemma~1, choose $0 < \eps< \eps_1 $ and let 
\begin{align} \label{e:def R} 
R_n(z, \zprime) & = \frac{T_n(z, \zprime) - \eps \pi(\zprime) } { 1-\eps} 
\end{align}
Then, $R_n(z, \zprime)$ is a one-step transition kernel with the following properties. 
\begin{lemma}
(a) 
\begin{align*} 
\sum_z \pi(z) R_n(z, \zprime) & = \pi(\zprime) \ . 
\end{align*}
(b) 
\begin{align*}
R_n(z, \zprime) & \leq Kg_0(z_n) + \eta \delta_z (\zprime) 
\end{align*}
where $0 < \eta < 1 $. \\
(c) 
\begin{align*}
\mid R_n(z, \zprime) - R_{n+1} (z, \zprime) \mid \ \leq \ K a_n 
\biggl \{  \biggl ( g_0(\zprime) + \frac{\pi(\zprime) } { \pi(z) } g_0(z) \biggr ) + \biggl ( 1 + \frac{g_0(z)} { \pi(z)} \biggr ) \delta_z(\zprime) \biggr \} 
\end{align*} 

(d) 
\begin{align*}
\sum_{z_{n-m+1}} \cdots \sum_{z_{n-1}} \prod_{k=1}^m R_{n-k} ( z_{n-k}, z_{n-k+1}) & \leq K g_0(z_n) + \eta^m \delta_{z_{n-m}}(z_n) 
\end{align*}

(e)~For $ 1 \leq l \leq j-1$ and $j=1, \dots, n $, 
\begin{align*}\sum_{z_{n-l-1} } \cdots \sum_{z_{n-j+1}} \pi(z_{n-j+1}) \prod_{k=1+1}^{j-1} R_{n-j} (z_{n-k}, z_{n-k+1}) & = \pi(z_{n-l}) 
\end{align*}

(f) For  $ j = 1, \dots, n$ and $l = 1, \dots, j-1$, 
\begin{align*}
\mid \sum_{z_{n-j+1}} \cdots \sum_{z_{n-1}} \pi(z_{n-j+1} ) \prod_{k=l+1}^{j-1}  R_{n-j}(z_{n-k}, z_{n-k+1}) & \\
\times \biggl ( R_{n-l} (z_{n-l}, z_{n-l+1}) - R_{n-j} (z_{n-l}, z_{n-l+1}) \biggr ) \prod_{k=1}^{l-1} R_{n-k}(z_{n-k}, z_{n-k+1})\mid & \ \leq \ Kg_0(z_n) a_{n-j}(j-l)^{-1} 
\end{align*}
\end{lemma}

\begin{proof} [Proof] (a) follows from \eqref{e:def R} and \eqref{e:T invar}. (b) follows from \eqref{e:def R}. (c) follows from \eqref{e:dimn adap delta} and \eqref{e:dimn adap nu}. (d) is true for $m = 1$ and is obtained in general by induction. (e)~follows from part~(a). (f)~follows from parts~(a) to (e). 
\end{proof}

\begin{proof}[Proof of Theorem 1] 
Let $\Delta_j , j=1, 2, \dots $ be an independent Bernoulli process such that $\Delta_j = 1$ with probability $\eps$ and $\Delta_j = 0 $ with probability $1-\eps$. From \eqref{e:def R}, 
$
T_n(z,\zprime) = (1-\eps) R_n(z, \zprime) + \eps \pi(\zprime) 
$
so that we can interpret $T_n(z,\zprime)$ as a mixture of transition kernels, such that $T_n(z,\zprime) = R_n(z,\zprime)$ if $\Delta_n = 0$ and $T_n(z,\zprime) = \pi(\zprime)$ if $\Delta_n = 1$. For $j = 1, \dots , n$, let $A_{n,j}$ be the event that $\Delta_{n-j+1} = 1, \Delta_k = 0, k=n-j+2, \dots , n$. Let $B_n$ be the event that $\Delta_j = 0$ for $j=1, \dots, n$. Then $\Pr(A_{n,j} ) = \eps(1-\eps)^{j-1} $ and $\Pr(B_n) = (1-\eps)^n$, and  
\begin{align*}
\Pr(Z_n = z_n) & = \sum_{j=1}^n \Pr(Z_n=z_n\mid A_{n,j} )\Pr(A_{n,j}) + \Pr(Z_n = z_n\mid B_n) \Pr(B_n) . 
\end{align*}
As in the proof of Theorem~1 in \cite{nott05}, we can write 
$\Pr(Z_n=z_n|A_{n,j}) = C_{0,j} + C_{1,j} + \cdots + C_{j-1,j}$, where 
\begin{align*}
C_{l,j} & = \sum_{z_0} \cdots \sum_{z_{n-1}} \Pr(Z_{0:n-j} = z_{0:n-j}) \pi(z_{n-j+1})
 \prod_{k=l+1}^{j-1} R_{n-j} (z_{n-k}, z_{n-k+1} ) \\
 &  \biggl ( R_{n-l} (z_{n-l},  z_{n-l+1} ) - R_{n-j} (z_{n-l}, z_{n-l+1})  \biggr ) \prod_{k=1}^{l-1} R_{n-k} (z_{n-k}, z_{n-k+1})  
\end{align*}
From part~(e) of Lemma~2,  $C_{0,j} = \pi(z_n)$ and by part (f) of Lemma~2,  
 $\mid C_{j,n}\mid \leq K g_0(z_n) a_{n-j}(j-l)^{-1}$ for $j >  1$. 
 Using a similar argument to that in \cite{nott05}, this implies that 
 \begin{align*}
 \biggl | \sum_{l=0}^{j-1} C_{l,j} \biggr | & \leq \pi(z_n) + K g_0(z_n) 
 (n-j)^{-r_1} j^{2} \ . 
\end{align*} 
Thus, 
\begin{align*}
& \sum_{j=1}^n \Pr(Z_n = z_n|A_{n,j} ) \Pr(A_{n,j}) = \pi(z_n) - (1-\eps)^n 
\pi(z_n) + \eta_n \quad \text{where} \\
& \mid \eta_n \mid  \ \leq \ K n^{-r_1} \sum_{j=1}^{n-1} \biggl ( 1 - \frac{j}{n} 
\biggr) ^{-r_1} j^{2} \eps(1-\eps)^{j-1} . 
\end{align*}
We also have that 
\begin{align*}
\Pr(Z_n = z_n\mid B_n) & = \sum_{z_0} \cdots \sum_{z_{n-1}} q_I(z_0) q_I(z_1) 
\prod_{k=1}^{n-1} R_{k}(z_k, z_{k+1}) \\
 &  \leq K g_0(z_n) + \eta^{n-1} q_I(z_n) \leq Kg_0(z_n) \ .  
 \end{align*}
 using Lemma~2~(c) and \eqref{e:domin}. 
Hence, 
 \begin{align} \label{e:prob ineq} 
 \mid \Pr(Z_n = z_n) - \pi(z_n) \mid \ \leq \ Kg_0(z_n) \biggl ( (1-\eps)^n  + n^{-r_1} \biggr )
\end{align}
The proof of Theorem~1 follows. 
\end{proof} 

The proof of Theorem~2 is similar to that in \cite{nott05} if we use \eqref{e:prob ineq}.

\newpage

\textheight  10.5in


\end{doublespace}
\end{document}